# A reusable metasurface template

*Jianxiong Li[1], Ping Yu[1], Shuang Zhang[2*], and Na Liu[3,4*]*

[1]Max Planck Institute for Intelligent Systems, Heisenbergstrasse 3, 70569 Stuttgart, Germany

[2]School of Physics & Astronomy, University of Birmingham, Birmingham B15 2TT, UK.

[3]2nd Physics Institute, University of Stuttgart, Pfaffenwaldring 57, 70569 Stuttgart, Germany

[4]Max Planck Institute for Solid State Research, Heisenbergstrasse 1, 70569 Stuttgart, Germany

Metasurfaces have revolutionized the design concepts for optical components, fostering an exciting field of flat optics. Thanks to the flat and ultrathin nature, metasurfaces possess unique advantages over conventional optical components, such as light weight, high compatibility, among others. To comply with potential applications, further research endeavors need to be exerted for advancing the reusability and practicality of metasurfaces. In this work, we demonstrate a reusable template approach to achieve optical multifunctionality with metasurfaces, utilizing both the geometric and propagation phases to shape light waves. In principle, such a metasurface template can be employed infinite times to enable a large variety of optical functions. As proof of concept experiments, we demonstrate metalensing, holography, and vortex beam shaping. Our work will leverage the high scalability aspects of metasurface devices for practical applications.

Metasurfaces bring forward a new class of optical devices, which can control light waves on flat and ultrathin platforms with exceptional flexibilities.[1-5] Different from conventional optics based on light propagation and refraction, metasurfaces utilize subwavelength structures, such as metallic and dielectric antennas to resonantly scatter light with the desired amplitude, phase, and polarization.[6-11] This capability to sculpt light waves with nanoscale accuracy has fostered a plethora of exciting optical functions and rapidly expanded the scope of potential applications. The further development of this field entails not only continuous endeavors to optimize the device performance[12-15] but also smart methodologies to enhance the device adaptability for practical applications and future commercialization.[16-23] Given the flat and ultrathin nature, metasurfaces already possess unique advantages, such as light weight, high compatibility, and ease of fabrication.[24-29] Generally, the implementation of metasurface devices follows the strategy of one structural design for one particular optical function. Although there have been a number of works on multifunctional and reconfigurable metasurfaces,[30-38] often such metasurfaces were engineered to exhibit only single optical functions in a dynamic fashion, for instance, focal length variations for dynamic focusing,[16, 19, 39, 40] refraction or reflection angle sweeping for beam steering,[33, 41, 42] among others. Endowing a single metasurface with miscellaneous optical functions still faces significant challenges and awaits more research endeavors, especially at visible frequencies.

In this work, we propose and demonstrate a metasurface template concept. With a single metasurface template, diverse optical functions are achievable at visible frequencies. The metasurface template utilizes both the geometric and propagation phases to shape light waves. The geometric phase is determined by the fix orientations of anisotropic plasmonic antennas based on the Pancharatnam−Berry (PB) phase. The propagation phase is introduced by a spatially patterned dielectric layer of subwavelength dimension, which are erasable and rewritable. In principle, the metasurface template can be employed infinite times to generate a large variety of optical functions. As proof of concept experiments, we demonstrate metalensing, holography, and vortex beam generation using the same metasurface template. We envision that the metasurface template can be both rewritable and reconfigurable, when a stimulus-responsive material is utilized to serve as the spatially patterned dielectric layer. This will largely enrich the applicability and practicality of the underlying concept.

Figure 1a shows the concept of the reusable metasurface template. It consists of gold nanorods arranged in a specific geometric phase profile on a quartz substrate. Such a metasurface then works as a template for incorporation of different propagation phase distributions to yield diverse optical functions including metalensing, holography, vortex beam shaping, among others. More specifically, as shown in Fig. 1b, four gold nanorods (pixels, $P_1$-$P_4$) with dimensions of 200 nm × 80 nm × 30 nm are oriented with angles of 0, $\pi/4$, $\pi/2$ and $3\pi/4$ in each unit cell (600 nm × 600 nm), generating PB phase delays of 0, $\pi/2$, $\pi$, and $3\pi/2$ accordingly. At normal incidence of circularly polarized light (633 nm), the intensity of the output light from the metasurface with cross-polarization is zero due to the destructive interference among the four antennas with different geometric phases ($\pi$ phase difference between $P_1$ and $P_3$ as well as between $P_2$ and $P_4$). This phenomenon is also revealed in the two-dimensional spatial frequency spectrum obtained through the Fourier transform of the phase profile as shown in Fig. 1c. Two spatial frequency components $F_1$ (1.055, 0) $\lambda^{-1}$ and $F_2$ (0, 1.055) $\lambda^{-1}$, corresponding to the first diffraction orders along the *x*- and *y*-directions, respectively, appear in the spatial frequency spectrum. They correspond to the excitations of evanescent waves along the *x*- and *y*-directions, respectively. When a homogeneous dielectric layer (*e.g.*, PMMA) of subwavelength thickness is fully covered on the metasurface, an additional phase △φ, resulting from the combination of light propagation through the PMMA layer and the resonance shift of the nanorod, is introduced to each pixel (see Supplementary Fig. S1). Nevertheless, there is no light output from the metasurface, as the relative phase differences among the pixels remain the same. If the PMMA layer on top of one pixel (*e.g.*, $P_3$) within a unit cell is removed through electron-beam lithography (EBL) followed by resist development, the phase difference between $P_1$ and $P_3$ is changed to $\pi$-$\triangle\varphi$, whereas the phase difference between $P_2$ and $P_4$ is still $\pi$. Two new spatial components $F_3$ (0, 0) $\lambda^{-1}$ and $F_4$ (1.055, 1.055) $\lambda^{-1}$, which correspond to the excitations of a propagating wave and an evanescent wave, respectively, appear in the spatial frequency spectrum (see Fig. 1c). Figure 1d shows the intensities of the four spatial frequency components ($F_1$-$F_4$) in dependence on $\triangle\varphi$, which in turn depends on the PMMA thickness. When $\triangle\varphi$ is tuned to $\pi$, the maximum intensity of the propagating wave ($F_3$) is achieved, owing to the zero phase difference between $P_1$ and $P_3$, which leads to constructive interference in the far field. In this case, the overall effect of the unit cell is to yield a phase delay of $\pi$. In other words, removals of the PMMA on top of $P_1$, $P_2$, $P_3$, and $P_4$ within a unit cell give rise to selective phase delays of 0, $\pi/2$, $\pi$, and $3\pi/2$ accordingly. As a result, a specific phase delay can be readily achieved from each unit cell on this four phase-level metasurface simply by making a PMMA opening on the corresponding pixel. After dissolving the PMMA by acetone, the phase information $\triangle\varphi$ is completely erased, so that the metasurface is restored to the initial state. Subsequently, a new PMMA layer can be patterned on the template, giving rise to a different phase profile. In this regard, the metasurface can serve as a rewritable template to yield a variety of optical functions. In the following, we will showcase the reusability of the metasurface template by demonstrating several representative optical functions, including metalensing, holography, and vortex beam generation.

We first investigate the realization of a metalens using the metasurface template. Figure 2a illustrates the four different phase delays achieved by introducing local PMMA openings (300 nm × 300 nm) on the corresponding pixels in a unit cell. Our numerical simulation shows that the performance of the metasurface is highly tolerant of the misalignment between the PMMA opening (Supplementary Fig. S2). Specifically, along different directions alignment deviations up to 120 nm do not lead to significant influences on the desired phase delay. In Fig. 2b, the required phase distribution for a metalens with a focal length $f$ of 750 μm is calculated at an operating wavelength of 633 nm, wherein the four phase delays are marked by squares of different colors. The focusing performance of the lens based on the metasurface template is highly dependent on $\triangle\varphi$. As shown in Fig. 2c, the maximum light intensity is achieved at the focal plane ($z$ = 750 μm), when $\triangle\varphi = \pi$. Figure 2d presents the simulated focusing effect of the metalens with $\triangle\varphi = \pi$. The light intensity profile at the focal plane forms an airy pattern as characterized by the red curve in Fig. 2e. Figure 2e also further confirms the intimate relation between the focusing performance and $\triangle\varphi$.

The experimental demonstration of the metalens is presented in Fig. 3a. The scanning electron microscopy (SEM) image of the metasurface is shown in the same figure. The PMMA openings at the selected pixel positions are clearly visualized in the different unit cells. The PMMA thickness is 270 nm. The incident light is right-handed circularly polarized (RCP) at 633 nm. The snapshot images reveal the beam shape evolution at the representative $z$-positions from the metalens. A pronounced focal point is clearly observed at $z$ = 752 μm, which is in a good agreement with the simulated results. In order to validate the rewritability of the template, the PMMA layer is removed by acetone. Subsequently, new PMMA openings are patterned on the template to acquire a phase profile for a cylindrical metalens with a focal length of 500 μm. The SEM image of the metasurface and optical characterizations of the cylindrical metalens are shown in Fig. 3b. The experimentally measured focal length is 497 μm. As a control experiment, a sample with only PMMA openings that are arranged in the same spatial distribution as that on the metalens in Fig. 3a is fabricated. The corresponding SEM image is shown in Fig. 3c. It is evident that no focusing effect is achieved, as the working principle involves the spatial variations of both the geometric phase and the propagation phase on the metasurface template.

To demonstrate the versatile applicability, the metasurface template is also applied in a reflection mode to reconstruct computer-generated holograms. In this case, the gold nanorods reside on a $SiO_2$/Si substrate. The PMMA openings are distributed on the template following a hologram phase profile calculated using Gerchberg-Saxton algorithm.[43] The SEM image of the metasurface is shown in Fig. 4a. Under normal illumination of RCP light at 633 nm, a high-quality holographic pattern of "sixteenth note" is observed at an off-axis angle of 11°. When the helicity of the incident light is changed from RCP to LCP, its centrosymmetrically inverted image appears. This indicates that the optical function enabled by the metasurface template essentially possesses a geometric phase nature. Upon the removal of PMMA, all the propagation phase information is erased and only the geometric phase information remains. No hologram is observable as shown in Fig. 4b. When another propagation phase information is introduced by patterning a new PMMA profile on the metasurface template, a holographic of "treble clef" is reconstructed as shown in Fig. 4c.

To further corroborate the generality of the metasurface template concept, we also demonstrate vortex beam generations in Fig. 5. Through multiple erasing and rewriting processes, vortex beams with different topological charges are formed using one single template. Figure 5a shows the experimental vortex beam intensity ($l = 1$, red curve) and simulated $\triangle\varphi$ (black curve) as a function of the PMMA thickness $t$. It reveals that the maximum light intensity is achieved when $t = 270$ nm. This indicates that the condition of $\triangle\varphi \approx \pi$ is fulfilled. PMMA openings with $t = 270$ nm are adopted in the experiment and the SEM image of a representative metasurface ($l = 1$) is shown in Fig. 5b. The schematic phase illustrations and measured light intensity profiles of the vortex beams with topological charges of $l = 1, 2, 3$ are presented in Figs. 5c and 5d, respectively. As shown in Fig. 5d, the intensity profiles exhibit typical doughnut shapes. The central dark area of the light beam becomes larger with the increase of the topological charge.[44] The non-uniform ring profile of the vortex beam results from the discontinued phase distribution (4 phase levels) of the metasurface template and nanofabrication imperfections. The optical efficiency of the metasurface template is 21% in this case. It can be improved through structural optimizations and/or implementing the design concept to dielectric metasurfaces.

In conclusion, we have demonstrated a metasurface template concept by combination of both the geometric and propagation phase controls. The propagation phase information is erasable and rewritable, enabling the reusability of the metasurface template. In principle, one metasurface template can be applied infinite times to yield miscellaneous optical functions, largely reducing the fabrication complexities. As proof-of-concept experiments, we have utilized the metasurface template to construct metalenses, computer-generated holograms, and vortex beams. Our design principle is generic and a variety of functional materials can be incorporated on the metasurface template. Apart from PMMA that is specific to electron-beam writing, phase-transition materials, such as vanadium dioxide and germanium-antimony-tellurium, especially those that are compatible with direct laser writing can be adopted to enable metasurface templates that are both rewritable and reconfigurable. In such cases, direct laser writing can locally trigger the refractive index changes of the phase-transition materials to generate the desired propagation phase distributions. Subsequently, the propagation phase information can be erased upon resetting of the phase-transition materials, for instance, through thermal treatments. Our work offers a viable route to create metasurface devices with reusability. It also stimulates inspiring concepts to design novel metasurfaces with practical features for real-world applications.

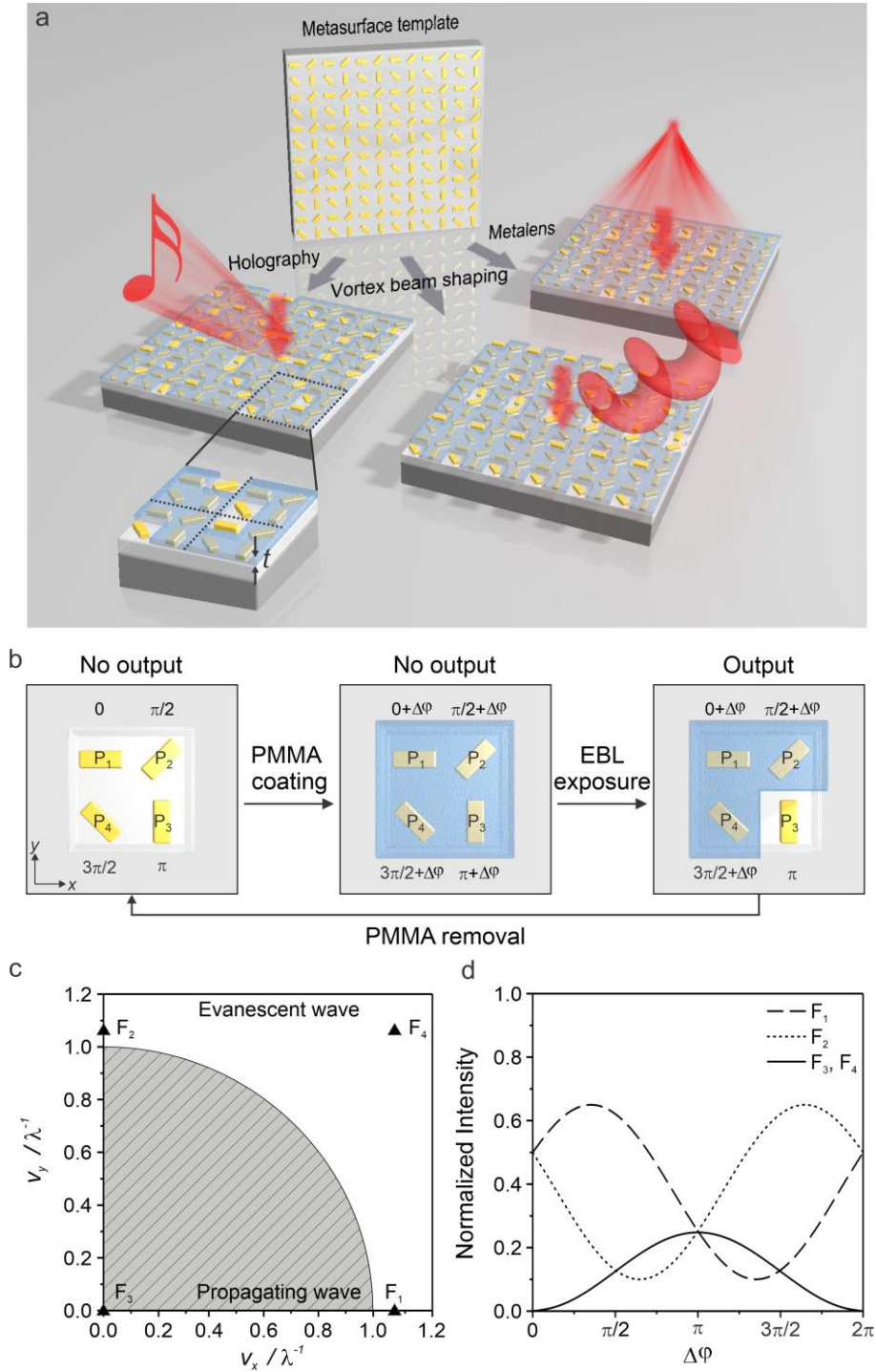

**Figure 1.** Working principle of the metasurface template. (a) Schematic of the reusable metasurface template. Gold nanorods are arranged in a specific geometric phase profile on the metasurface, which works as a template for incorporation of different propagation phase distributions to yield diverse optical functions, including metalensing, computer-generated holography, vortex beam shaping, among others. (b) Phase delays generated from the four

pixels (P$_1$-P$_4$) within a unit cell without PMMA, with full PMMA coverage, and with a selective PMMA opening. (c) Two-dimensional spatial frequency spectrum calculated from the discrete phase distribution profile of the metasurface template. The circle of $v_x^2 + v_y^2 = \lambda^{-2}$ separates the zones of output propagating (grey) and evanescent (white) waves. F$_1$-F$_4$ correspond to the four spatial frequency components. (d) Normalized intensities of F$_1$-F$_4$ in dependence on $\Delta\varphi$. F$_3$ and F$_4$ follow the same curve, which is presented by the solid line.

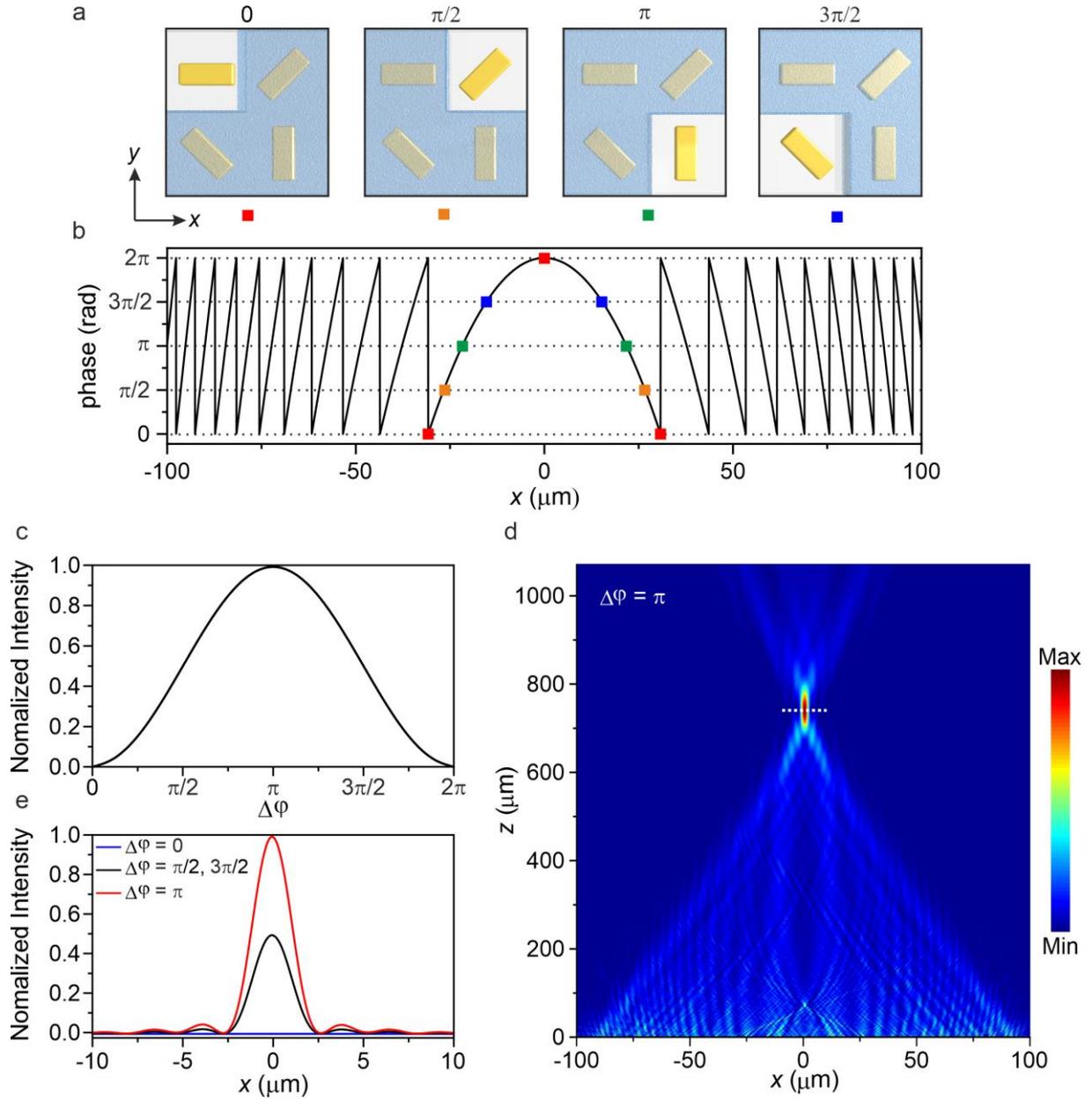

**Figure 2.** Simulation of the metalenses based on the metasurface template. (a) Four different phase delays 0, π/2, π and 3π/2, achieved by local removals of PMMA on the corresponding pixels in a unit cell. (b) Calculated phase profile of the metalens with a focal length of 750 μm. The four phase delays in (a) are marked with different colors. (c) Light intensity at the focal plane ($z$ = 750 μm) as a function of Δφ. (d) Simulated focusing effect of the metalens based on the metasurface template, when Δφ = π. (e) Light intensity profiles along the $x$-direction at the focal plane (white line in d), when Δφ = 0, π/2, π, and 3π/2.

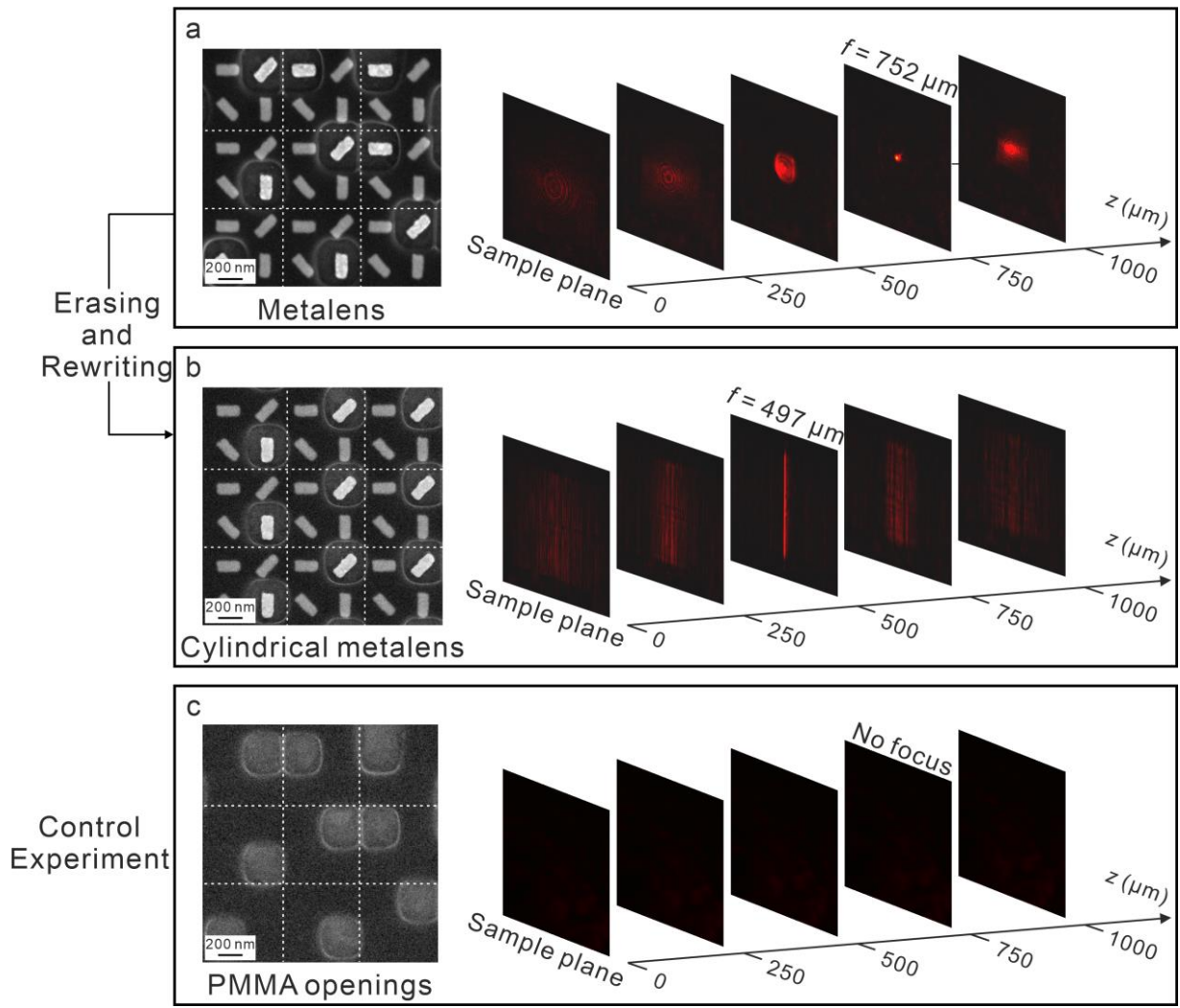

**Figure 3.** Metasurface template for lensing. SEM images and representative intensity snapshots in the *x*−*y* plane at different z-distances from (a) metalens with a focal length of 752 μm, (b) cylindrical metalens with a focal length of 497 μm and (c) control sample with only PMMA openings. The control sample has PMMA openings arranged in the same spatial distribution as that on the metalens in (a).

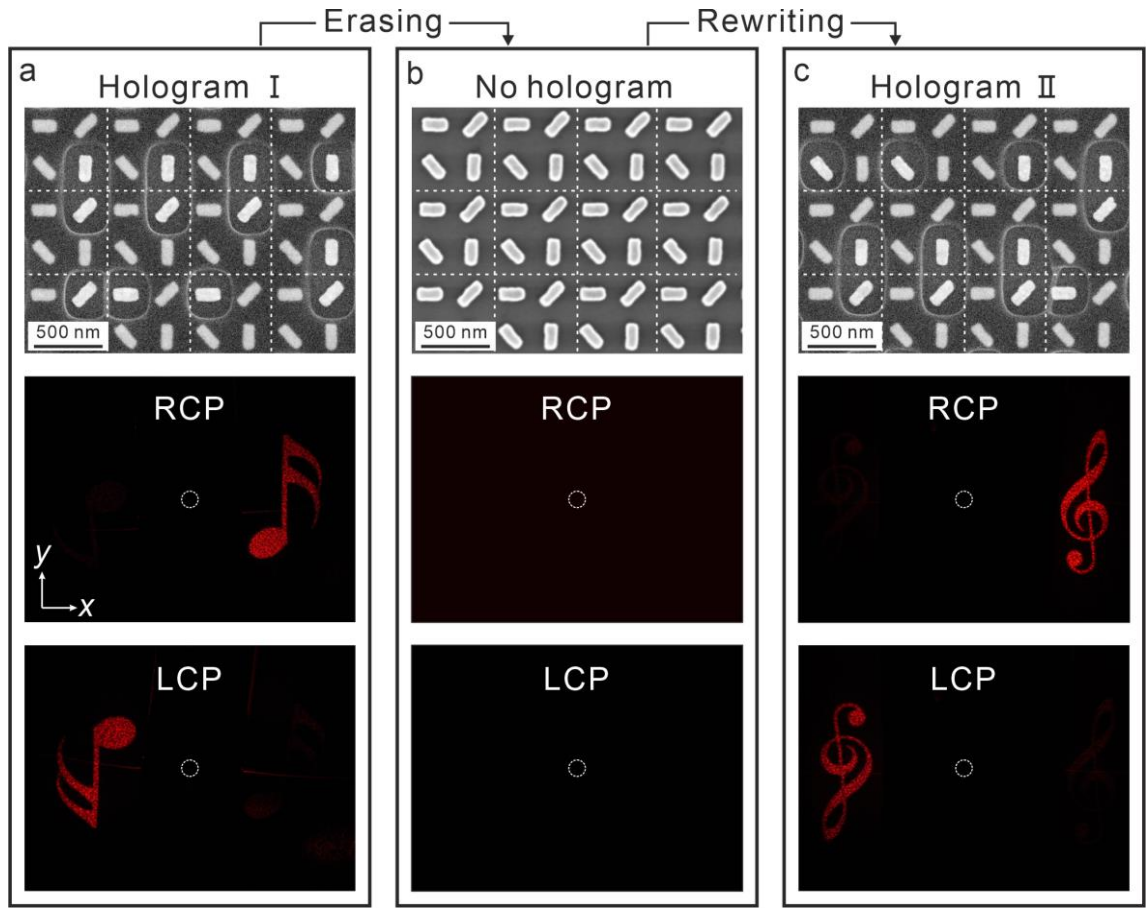

**Figure 4.** Metasurface template for holography. SEM images and representative snapshots of the holographic images (a) 'sixteenth note', (b) blank, and (c) 'treble clef' under RCP and LCP incident light. The central point in each plot is indicated by a white circle, highlighting the location of the zero-order reflected light.

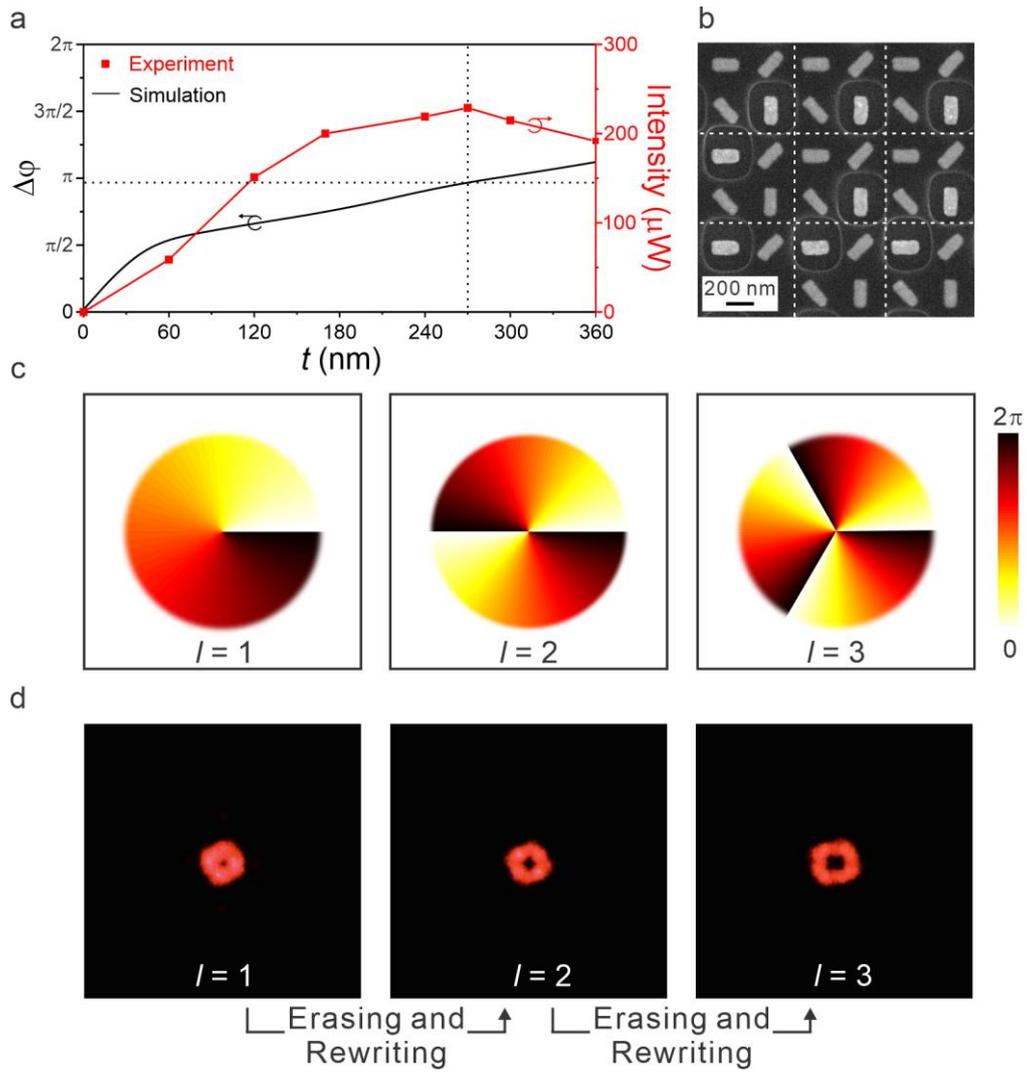

**Figure 5.** Metasurface template for vortex beam shaping. (a) Experimental vortex beam intensity ($l = 1$, red curve) and simulated $\triangle\varphi$ (black curve) as a function of the PMMA thickness $t$. (b) SEM image of the metasurface template for vortex beam generation ($l = 1$). (c) Schematic phase illustrations and measured light intensity profiles of the vortex beams with topological charges of $l = 1, 2, 3$.

## ASSOCIATED CONTENT

**Supporting Information**.

The following files are available free of charge.

Supporting information (word)

## AUTHOR INFORMATION

**Corresponding Author**

*Email: s.zhang@bham.ac.uk

*Email: na.liu@pi2.uni-stuttgart.de

**Author Contributions**

J.X.L. and N.L. conceived the project. P.Y. and S. Z. provided crucial suggestions to the main concept of the project. J.X.L. performed the experiments and theoretical calculations. All authors discussed the results, analyzed the data, and commented on the manuscript.

**Notes**

The authors declare no competing financial interest.

## ACKNOWLEDGMENT

This project was supported by the European Research Council (ERC Dynamic Nano and ERC Topological) grants.

ToC Figure:

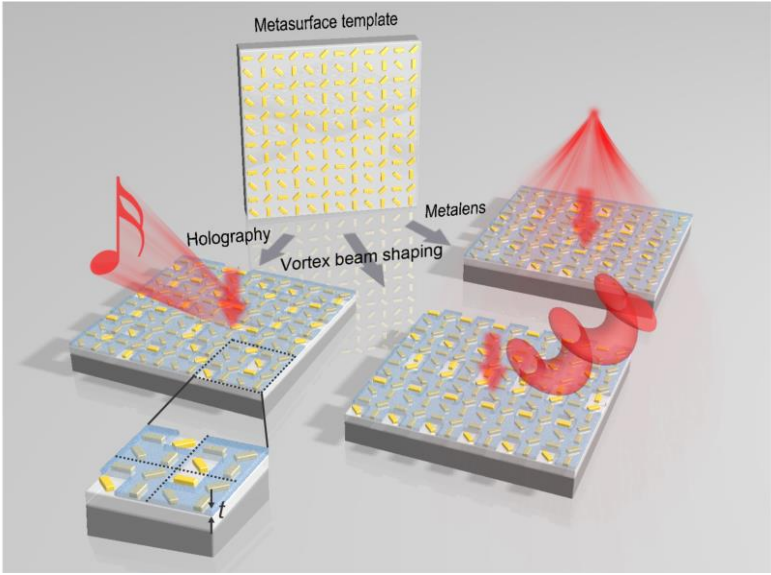